\documentclass{piparticle-final}
\usepackage[utf8]{inputenc}
\usepackage{graphicx}
\usepackage{amsmath}
\usepackage{amssymb}

\usepackage{cite} 
\usepackage{epstopdf}

\begin{document}

\volume{7}               
\articlenumber{070012}   
\journalyear{2015}       
\editor{C. A. Condat, G. J. Sibona}   
\reviewers{A. De Luca, Laboratoire de Physique Theorique, \\ \mbox{} \hspace{3.62cm} ENS \& Institut Philippe Meyer, Paris, France.}  
\received{20 November 2014}     
\accepted{29 June 2015}   
\runningauthor{P. R. Zangara \itshape{et al.}}  
\doi{070012}         

\title{Role of energy uncertainties in ergodicity breaking induced by competing interactions and disorder. A dynamical assessment through the Loschmidt echo.}

\author{Pablo R. Zangara,\cite{inst1,inst2}\thanks{Email: zangara@famaf.unc.edu.ar}\hspace{0.5em}
	Patricia R. Levstein,\cite{inst1,inst2}\hspace{0.5em}
	Horacio M. Pastawski\cite{inst1,inst2}\thanks{Email: horacio@famaf.unc.edu.ar}
	}

\pipabstract{
A local excitation in a quantum many-particle system evolves
deterministically. A time-reversal procedure, involving the invertion
of the signs of every energy and interaction, should produce an excitation
revival: the Loschmidt echo (LE). If somewhat imperfect, only a fraction of
the excitation will refocus. We use such a procedure to show how non-inverted
weak disorder and interactions, when assisted by the natural reversible
dynamics, fully degrade the LE. These perturbations enhance diffusion and
evenly distribute the excitation throughout the system. Such a dynamical
paradigm, called ergodicity, breaks down when either the disorder or the
interactions are too strong. These extreme regimes give rise to the well
known Anderson localization and Mott insulating phases, where quantum
diffusion becomes restricted. Accordingly, regardless of the kinetic energy
terms, the excitation remains mainly localized and out-of-equilibrium, and
the system behaves non-ergodically. The LE constitutes a fair dynamical
witness for the whole phase diagram since it evidences a surprising
topography in which ergodic and non-ergodic phases interpenetrate each
other. Furthermore, we provide an estimation for the critical lines
separating the ergodic and non-ergodic phases around the Mott and Anderson
transitions. The energy uncertainties introduced by disorder and interaction
shift these thresholds towards stronger perturbations. Remarkably, the
estimations of the critical lines are in good agreement with the phase
diagram derived from the LE dynamics.
}

\maketitle

\blfootnote{
\begin{theaffiliation}{99}
	\institution{inst1} Instituto de F\'{i}sica Enrique Gaviola (CONICET-UNC), Argentina.
	\institution{inst2} Facultad de Matem\'{a}tica, Astronom\'{i}a y F\'{i}sica, Universidad Nacional de C\'{o}rdoba, 5000 C\'{o}rdoba, Argentina.
\end{theaffiliation}
}

\section{Introduction}

According to Classical Mechanics, a system composed by $N$ particles in $d$
dimensions is described as a point $X$ in a ($2dN$)-dimensional phase space.
If the system is conservative, the energy is the primary conserved quantity,
and the phase space is restricted to a hypersurface $\mathcal{S}$ of $2dN-1$
dimensions usually called \textit{energy shell}. Fully integrable systems
are further constrained, and their solutions turn out to be regular and
non-dense periodic orbits contained in $\mathcal{S}$. If integrability is
broken, the orbits become irregular and cover $\mathcal{S}$ densely. This
means that the actual trajectory $X(t)$ will uniformly visit every
configuration within $\mathcal{S}$, provided that enough time has elapsed.
This last observation embodies the concept of \textit{ergodicity}:\ an
observable can be equivalently evaluated by averaging it for different
configurations in $\mathcal{S}$ or by its time-average along a single
trajectory $X(t)$. In such a sense, ergodicity sets the equivalence between
the Gibbs' description in terms of ensembles and Boltzmann kinetic approach
to Thermostatistics. Therefore, the ergodic hypothesis has become the
cornerstone of Classical Statistical Mechanics \cite%
{LebowitzPhysToday,LebowitzRMP1999}.

Almost 60 years ago, E. Fermi, J. Pasta and S. Ulam (FPU) \cite{FPU} tried
to study when and how the integrability breakdown can lead to an ergodic
behavior within a deterministic evolution. They considered a string of
harmonic oscillators perturbed by anharmonic forces in order to verify that
these non-linearities can lead to energy equipartition as a manifestation of
ergodicity. Even though Ulam himself stated \textquotedblleft \textit{The
motivation then was to observe the rates of mixing and thermalization...}%
\textquotedblright\ \cite{fermi_book}, the results were not the expected
ones: \textquotedblleft thermalization\textquotedblright\ dynamics did not
show up at all. Nowadays, their striking results are well understood in
terms of the theory of chaos \cite{izraChaosReview}. In this context, chaos is defined as an exponential sensitivity to changes in
the initial condition. In fact, the onset of dynamical chaos \cite%
{chirikov1,chirikov2} can be identified with the transition from non-ergodic
to ergodic behavior. Therefore, within classical physics, the emergence of
ergodicity can be satisfactorily explained \cite{Zaslavsky}.

The previous physical picture cannot be directly extended to Quantum
Mechanics. Indeed, \textit{any} closed quantum system involves a discrete
energy spectrum and evolves quasi-periodically in the Hilbert space, which
becomes the quantum analogue to the classical phase space. Nevertheless,
thermalization and ergodicity in isolated quantum systems could still be
defined for a set of relevant observables \cite{LebowitzQET,vonNeumannQET}.
Since the sensitivity to initial conditions does not apply to quantum
systems, the quantum signature of dynamical chaos had to be found as an
instability of an evolution towards perturbations in the Hamiltonian \cite{jalpa}. Because this definition encompasses the classical one,
it builds a bridge between classical and quantum chaos. Moreover, since it also
implies an instability towards perturbations in a time reversal procedure,
it can be experimentally evaluated  \cite{patricia98} as the amount of
excitation recovered or Loschmidt echo (LE) \cite{scholarpedia,Jacquod}.
Such a revival is degraded by the presence of uncontrolled environmental
degrees of freedom as in the usual picture of decoherence for open quantum
systems \cite{Zangara2012}. Strikingly, in closed systems with enough
internal complexity, even simple perturbations seem to degrade the LE in a
time scale given by the reverted dynamics, revealing how a mixing dynamics
drives irreversibility \cite{usaj-physicaA,MolPhys}.

Within the last years, a new generation of experiments on relaxation and
equilibration dynamics of (almost perfectly) closed quantum many-particle
systems has become accessible employing optical lattices loaded with cold
atoms \cite{CradleNature2006,BlochNatPhys2012}. These became the driving
force behind the recent theoretical efforts to grasp quantum thermalization 
\cite{polkovnikovRMP}.

Attempting a step beyond the FPU problem, the current aim is to study simple
quantum models that could go parametrically from an ergodic to a non-ergodic
quantum dynamics. Moreover, a fundamental question is whether such a
transition occurs as a smooth crossover or a sharp threshold.  A promising
candidate for these studies would be a system showing Many-Body Localization
(MBL) \cite{altshuler2006,Altshuler2010}. This dynamical phenomenon occurs
when an excitation in a disordered quantum system evolves in presence of
interactions. In fact, the MBL is a quantum dynamical phase transition
between extended and localized many-body states that results from the
competition between interactions \cite{MottRMP} and Anderson disorder \cite{AndersonRMP-1978}. If the many-body states are extended, then one may
expect that the system is ergodic enough to behave as its own heat bath. In
such case, single energy eigenstates would yield expectation values for
few-body observables that coincide with those evaluated in the
microcanonical thermal ensemble \cite{popescuNATURE2006,rigolNATURE2008}.
Quite on the contrary, if the many-body states are localized,  any initial
out-of-equilibrium condition would remain almost frozen. In this case,
self-thermalization is precluded. Therefore, the MBL would evidence the
sought threshold between ergodic and non-ergodic behavior.

In this article, we address the competition between interactions and
disorder in a one-dimensional (1D) spin system. It is already known that such
models evidence the MBL transition, at least for particular parametric
regimes \cite{oganesyanhuse2007,prosen2008,palhuse2010,pollmann2012,DeLuca2012}. Our
approach to tackle this problem involves the evaluation of the LE, here
defined as the amount of a local excitation recovered after an imperfect
time reversal procedure. This involves the inversion of the sign of the
kinetic energy terms in the many-spin Hamiltonian \cite{Zangara2012}.
Moreover, the LE is evaluated as an autocorrelation function that could
become a suitable order parameter \cite{Refael2013}. Thus, the LE is a
natural observable that allows us to identify when the ergodicity of an
excitation dynamics is broken as interactions and disorder become strong
enough. When weak, these \textquotedblleft perturbations\textquotedblright\
favor the excitation spreading, but limit LE recovery as they are not
reversed.

From the actual LE time-dependence in the \textit{infinite}-\textit{temperature} regime, we extract a dynamical phase diagram that shows a
non-trivial interplay between interactions and disorder. The \textit{near-zero temperature} regime has already been addressed in the literature 
\cite{Giamarchi1988,Fisher1992} and there are conjectures about the global
topography of the phase diagram \cite{Kimball1981}. In analogy with this
last case, we address the nature of two critical lines that separate the
ergodic phase from two different non-ergodic phases: the Mott insulator and
the MBL phase. The appearance of either, Mott insulator and MBL phases, can
be well estimated in terms of the relevant energy scales. Thus, in order to
evaluate an estimation of the critical lines, we compute the energy
uncertainties that weak disorder and weak interactions would impose on the \
states involved in the Mott transition and on the MBL transition,
respectively. Quite remarkably, these estimations show a good agreement with
the dynamical LE diagram. Our approach allows the identification of ergodic
and non-ergodic phases whose non-trivial structure may guide future
theoretical and experimental investigations.

\section{Loschmidt echo formulation}

We consider a 1D spin system that evolves according to a
Hamiltonian $\hat{H}=\hat{H}_{0}+\hat{\Sigma}$. Here, $\hat{H}_{0}$ stands
for a nearest neighbors $XY\footnote{Notice that, in the recent literature of strongly-correlated systems within
the condensed matter community, the notation $XX$ is employed instead of $XY$.}$ Hamiltonian \cite{mesoECO-exp,madi-ernst1997}:

\begin{eqnarray}
\hat{H}_{0} &=&\sum_{i=1}^{N}J\left[
S_{i}^{x}S_{i+1}^{x}+S_{i}^{y}S_{i+1}^{y}\right]  \label{H1} \\
&=&\sum_{i=1}^{N}\tfrac{1}{2}J\left[
S_{i}^{-}S_{i+1}^{+}+S_{i}^{+}S_{i+1}^{-}\right] ,
\end{eqnarray}%
which because of the periodic boundary conditions can be thought as arranged
in a ring. Here, unless explicitly stated, $N=12$. Notice that $\hat{H}_{0}$
can be mapped into two independent non-interacting fermion systems by the
Wigner-Jordan transformation \cite{danieli-CPL2004}. Therefore, it encloses
fully integrable single-particle dynamics.

The integrability of the model is broken by the Ising interactions and the
on-site disorder enclosed in $\hat{\Sigma}$,

\begin{equation}
\hat{\Sigma}=\sum_{i=1}^{N}\Delta
S_{i}^{z}S_{i+1}^{z}+\sum_{i=1}^{N}h_{i}S_{i}^{z},  \label{H2}
\end{equation}%
where $\Delta $ is the magnitude of the homogeneous interaction and $h_{i}$
are randomly distributed fields in the range $[-W,W]$. In order to enable
the comparison with the standard Anderson localization literature, we stress
that $W$ here is half of the standard strength commonly used for the
Anderson disorder \cite{MacKinnon93}.

The initial out-of-equilibrium condition is given by an \textit{infinite-temperature} state in which a local excitation (polarization) is
injected at site $1$: 
\begin{equation}
\left\vert \Psi _{neq}\right\rangle =\left\vert \uparrow _{1}\right\rangle
\otimes \left\{ \sum\limits_{r=1}^{2^{N-1}}\frac{1}{\sqrt{2^{N-1}}}e^{%
\mathrm{i}\varphi _{r}}\text{\ }\left\vert \beta _{r}\right\rangle \right\} ,
\label{neqstate}
\end{equation}%
where $\varphi _{r}^{{}}$ is a random phase and $\left\{ \left\vert \beta
_{r}\right\rangle \right\} $ are state vectors in the computational Ising
basis of the $N-1$ remaining spins. The state defined in Eq. (\ref{neqstate})
is a random superposition over the whole Hilbert space, and can successfully
mimic the dynamics of ensemble calculations \cite{Alv-parallelism}.
Additionally, notice that $\hat{\Sigma}$ perturbs the quantum phase of
each of the Ising states participating in the superposition.

Two evolution operators are built from the Hamiltonian operators \ref{H1}
and \ref{H2}, according to the relative sign between $\hat{H}_{0}$ and $\hat{%
\Sigma}$. These are $\hat{U}_{+}^{{}}(t_{R})=\exp [-\frac{\mathrm{i}}{\hbar }%
(\hat{H}_{0}^{{}}+\hat{\Sigma})t_{R}]$ and $\hat{U}_{-}^{{}}(t_{R})=\exp [-%
\frac{\mathrm{i}}{\hbar }(-\hat{H}_{0}^{{}}+\hat{\Sigma})t_{R}]$. In this
scenario, the LE is defined as the revival of the local polarization at site 
$1$:

\begin{align}
&M_{1,1}(2t)\notag\\
&=2\left\langle \Psi _{neq}\right\vert \hat{U}_{+}^{\dag }(t)\hat{U%
}_{-}^{\dag }(t)\hat{S}_{1}^{z}\hat{U}_{-}^{{}}(t)\hat{U}_{+}^{{}}(t)\left%
\vert \Psi _{neq}\right\rangle .  \label{eco}
\end{align}%
It is important to stress that Eq. (\ref{eco}) constitutes, at least in
principle, an actual experimental observable of the kind evaluated since the
early LE experiments \cite{patricia98,usaj-physicaA,MolPhys}, see also Ref. 
\cite{Ernst1992}. Moreover, both the local excitation and detection are well
established techniques within solid-state NMR \cite{Cappellaro2014_review}.
Nevertheless, changing the signs of specific Hamiltonian terms results in a
more subtle task. While the standard dipole-dipole interaction can be
reverted \cite{Rhim1971}, the planar $XY$ interaction requires much more
sophisticated pulse sequences, even for its forward implementation \cite%
{madi-ernst1997}. In particular, the mapping of the $XY$ interaction into a
double-quantum Hamiltonian, strictly valid in 1D systems \cite{elenaMQC},
could provide a novel approach to the problem of localization as it recently
did for 3D systems \cite{Alvarez2014experimental}. This could become a
pathway towards an experimental realization much related to the problem
considered here.

In order to analyze the ergodicity of our observable, we evaluate the mean
LE, $\bar{M}_{1,1}$:%
\begin{equation}
\bar{M}_{1,1}(T)=\frac{1}{T}\int_{0}^{T}M_{1,1}(t)dt.  \label{eco_medio}
\end{equation}

The standard analysis of localization implies the computation of $%
\lim_{T\rightarrow \infty }$ $\bar{M}_{1,1}(T)$. However, since the LE is
evaluated within a finite system, dynamical recurrences known as Mesoscopic
echoes show up at the (single-particle) Heisenberg time $T_{H}$ of $\hat{H}%
_{0}^{{}}$\textbf{.} As extensively discussed in Refs. \cite{mesoECO-PRL1995,mesoECO-exp,danieli-CPL2004}, $T_{H}$ can be estimated as

\begin{equation}
T_{H}\sim 2\sqrt{2}N\frac{\hbar }{J}.  \label{Theisenberg}
\end{equation}%
This estimation can be interpreted as the time needed by a local excitation
to wind around a ring of length $L=N\times a$ at an average speed $v_{M}/%
\sqrt{2}$, with a maximum group velocity $v_{M}=a\times \tfrac{1}{2}J/\hbar $%
. Since these recurrences are spurious to our analysis of the limiting case $%
N\rightarrow \infty $, we restrict our analysis to $T<T_{H}$.

Let us provide some specific details that should allow the reproduction of
our numerical computation. We evaluate Eq. (\ref{eco_medio}) ranging both $%
\Delta $ and $W$ within the interval $[0,5J]$, considering increments of $%
0.2J$ in both magnitudes. The relevant parameter regions were explored in
more detail by employing steps of $0.1J$. For each parameter set, $10$
realizations of disorder were averaged, each of them with second moment $%
\left\langle h^{2}\right\rangle =W^{2}/3$. For the non-interacting case $%
\Delta =0$, i.e., pure Anderson disorder, we computed $500$ disorder
realizations. With the purpose of keeping the statistical fluctuations
negligibly small, an extra average over $10$ realizations of $\left\vert
\Psi _{neq}\right\rangle $ is performed by tossing the phases $\varphi
_{r}^{{}}$ in the whole $[0,2\pi )$ range.

\section{A dynamical phase diagram}

\begin{figure}[ht]
\begin{center}
\includegraphics[width=0.99\columnwidth]{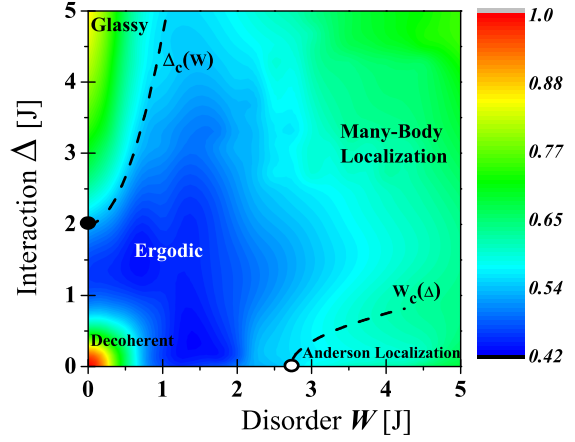}
\end{center}
\caption{Dynamical phase diagram: $\bar{M}_{1,1}(T)$ level plot at $%
T=12\hbar /J$ as a function of the interaction strength $\Delta $ and
disorder $W$.}
\label{Fig_mapa}
\end{figure}

Figure \ref{Fig_mapa} displays the dynamical phase diagram for the LE. It is
given by a level plot of $\bar{M}_{1,1}$ at $T=12\hbar /J$, as function of
the interaction $\Delta $ and disorder strength $W$. Within the diagram,
five dynamical regions are identified according to the predominant
mechanism: \textit{decoherent}, \textit{ergodic}, \textit{glassy}, \textit{%
Anderson localization}, and \textit{Many-Body Localization}.

If both $\Delta $ and $W$ are weak, the system is almost reversible, the
dynamics is controlled by single particle propagations and therefore $\bar{M}%
_{1,1}$ remains near $1$. This means that despite of the slight phase
perturbations, the local excitation can be driven back by the reversal of $%
\hat{H}_{0}$. Thus, the parametric region at the bottom left corner may be
associated with \textit{decoherence}, i.e., a sort of spin wave behavior
weakly perturbed by the imperfect control of the internal degrees of freedom 
\cite{madi-ernst1997,Fernandez2015}.

If either $\Delta $ or $W$ are further increased, the propagation of a local
excitation, ruled by $\hat{H}_{0}$, suffers  the effects of  $\hat{\Sigma}$
as multiple scattering events with the disordered potential and with other
spins. Thus, the excitation enters in a diffusive regime where it rapidly
spreads all over the spin system. As these scattering processes cannot be
undone by the reversal procedure, the spreading becomes irreversible.
Consistently, this bluish region is associated with an \textit{ergodic}
behavior for the polarization. In this regime, the polarization becomes
evenly distributed within the system, i.e., $2\left\langle \hat{S}%
_{j}^{z}\right\rangle =1/N$ for all $j$. The ideal limit $\bar{M}%
_{1,1}(T\rightarrow \infty )\rightarrow 1/N$ is verified up to a numerical
offset that comes from the transient decay of the LE.

If $W=0$, a strong increase in $\Delta $ leads to a predominance of the
Ising interaction, which freezes the polarization dynamics. Since the
quantum diffusion induced by $\hat{H}_{0}$ results drastically constrained, $%
\bar{M}_{1,1}$ remains trivially high. We interpret such behavior as a 
\textit{glassy} dynamics with long relaxation times. In fact, this sort of
localization corresponds to the Mott insulating phase of an impurity band 
\cite{MottRMP}. Additionally, the color contrast around $\Delta \gtrsim 2J$
suggests that the glassy-ergodic transition remains abrupt even for nonzero
disorder ($W\lesssim 1.0J$). This indicates a parameter region where the
interaction-disorder competition leads to a sharp transition between the
glassy and the ergodic phases. However, since transient phenomena become
very slow, a reliable finite size scaling of this regime would require
excessively long times to capture how a vitreous dynamics is affected by
disorder.

A dimensional argument provides a hint on the nature of the critical line
that separates the ergodic and glassy phases. In fact, the Mott transition
typically occurs when the interaction strength $\Delta $ is comparable to
the bandwidth $B=2J$. Such a particular interaction strength is singled out
in Fig. \ref{Fig_mapa} by a full black circle. Adding a weak disorder
introduces an energy uncertainty $\delta E$ on the energy levels that would
widen $B$. In order to estimate it, we resort to its corresponding time
scale $\tau $, which in turn can be evaluated according to the Fermi golden
rule (FGR). With such a purpose, we consider a \textit{localized} excitation
that can \textquotedblleft escape\textquotedblright\ either to its right or
to its left side, where two semi-infinite linear chains are symmetrically
coupled. Then,

\begin{equation}
\frac{1}{\tau }=2\frac{2\pi }{\hbar }\left( \frac{W^{2}}{3}\right)
N_{1}(\varepsilon ).  \label{FGR_W}
\end{equation}%
Here, as stated above, $W^{2}/3$ stands for second moment of the disorder
distribution. The factor $2$ accounts for the two alternative decays (right
and left). Additionally, $N_{1}(\varepsilon )$ is the Local Density of
States (LDoS) of a semi-infinite linear chain with hopping element $J/2$,

\begin{equation}
N_{1}(\varepsilon )=\frac{2}{\pi J}\sqrt{1-\left( \frac{\varepsilon }{J}%
\right) ^{2}}.  \label{LDoS_semiinfinito}
\end{equation}%
The energy levels acquire a Lorentzian broadening which, evaluated at the
spectral center $\varepsilon =0$, results

\begin{equation}
\delta E=\frac{\hbar }{2\tau }=\left. \frac{4}{3}\frac{W^{2}}{J}\sqrt{%
1-\left( \frac{\varepsilon }{J}\right) ^{2}} \right\vert _{\varepsilon =0}=\frac{4%
}{3}\frac{W^{2}}{J}.  \label{EnergyUncertainty_W}
\end{equation}%
Since half of the states lie beyond the range $B+2\delta E$, one may attempt
an estimation of the critical line for the Mott transition as,

\begin{equation}
\Delta _{c}(W)\sim B+2\delta E\sim 2J+\frac{8}{3}\frac{W^{2}}{J},
\label{delta_critico}
\end{equation}%
which is displayed in Fig. \ref{Fig_mapa} as a dashed line.

A similar functional dependence as the one discussed here for the
glassy-ergodic interphase was conjectured by Kimball for the interacting
ground state diagram \cite{Kimball1981}. Additionally, it is worthy to
mention that a naive expectation about the morphology of the phase diagram
with two competing magnitudes would be a semi-circular shape. This is
precisely the case of magnetic field and temperature as in the phase diagram
of a type I superconductor. Thus, one of the highly non-trivial implications
of the reentrance of the ergodic phase at large $\Delta $ in our diagram is
to debunk such an expectation.

If $\Delta =0$, the picture for $W>0$ is the standard \textit{Anderson
Localization} problem. Here, a reliable estimate of the localization length
is only possible when it is smaller than the finite size of the system. Not
being this the case of very weak disorder, the LE degrades smoothly as a
function of time with a dynamics that cannot be distinguished from a
diffusive one. When the disorder is strong enough, the localization length
becomes comparable with the lattice size and thus the initial local
excitation does not spread significantly.

Strictly speaking, while disordered 1D systems are always localized, there
are two mechanisms contributing to localization. One of them is the
\textquotedblleft strong localization\textquotedblright , i.e., the
convergence, term by term, of a perturbation theory for the local Green's
function. The other is the \textquotedblleft weak
localization\textquotedblright , originated in the interferences between
long perturbation pathways. This last one was an idea conceptually difficult to
grasp, both theoretically and numerically, until the appearance of the
scaling theory of conductance by the \textquotedblleft gang of
four\textquotedblright\ \cite{GangFour_PRL1979}. While weak localization is
particularly relevant in 1D and 2D systems, when these have a finite size
the dynamics remains diffusive and thus closely assimilable to an ergodic
one. In our problem, as soon as $\Delta \gtrsim 0$, the many-body interaction
increases the effective dimensionality of the available Hilbert space, and
thus it competes with the Anderson localization. Regardless of the precise
behavior near $\Delta =0$, such an interplay between $\Delta $ and $W$
is the responsible for the onset of a localization transition at some $%
W_{c}(\Delta )>0$, much as in a high dimensional lattice.

The ergodic-localized MBL transition can be observed when increasing $W$ for
a fixed $\Delta >0$. In particular, we notice that localization by disorder
is weakened when $1.0J\lesssim \Delta \lesssim 2.0J$, since the ergodic
region seems to unfold for larger $W$. Again, since we consider a \textit{%
finite} system, our observable describes a \textit{smooth crossover} from
the ergodic to a localized phase, where the excitation does not diffuse
considerably. In fact, this corresponds to the actual MBL phase transition 
\cite{oganesyanhuse2007,prosen2008,palhuse2010,pollmann2012,DeLuca2012},
which is genuinely sharp in the thermodynamic limit.

According to Eq. (\ref{Theisenberg}), $T\lesssim T_{H}\propto N$, and thus
increasing $N$ in our simulations (e.g., 10, 12 and 14) enables an
integration over a larger time $T$. In fact, when $\Delta \sim 1.0J$, we
verified that both sides of the MBL transition $\bar{M}_{1,1}$\ behave as
expected from physical grounds. Indeed, in the ergodic phase it has the
asymptotic behavior $\bar{M}_{1,1}\sim 1/N$, while in the localized phase of strong 
$W$ it saturates at \ $\bar{M}_{1,1}\sim 1/\lambda $, regardless of $N$. The
compatibility with a finite size scaling analysis is confirmed by the fact
that $\partial \bar{M}_{1,1}/\partial W$\ increases with $N$. However, our
accessible range for $N$ is not complete enough to provide for a scaling of $%
\bar{M}_{1,1}(T)$ that could yield precise critical values for the MBL
transition.

In analogy to the case of the Mott transition, a dimensional argument can be
performed to estimate the critical line $W_{c}(\Delta )$. Even though there
is no actual phase transition in the 1D non-interacting case\ $\Delta =0$,
as the interactions appear we expect them to break down the 1D constrains.
Thus, we consider as a singular point the high dimensional estimate that
occurs when the disorder strength is comparable to the bandwidth \cite%
{Ziman1969}

\begin{equation}
\left. W_{c}(\Delta )\right\vert _{\Delta =0}=(e/2)B.  \label{w_critico0}
\end{equation}%
The particular disorder strength in Eq. (\ref{w_critico0}) is indicated in
Fig. \ref{Fig_mapa} as an open circle, since it does not correspond to an
actual critical point of the 1D problem. Again, adding interactions would
introduce an energy uncertainty that widens the band accordingly. In this
case, the uncertainty $\delta E$ is associated to the lifetime introduced by
the Ising interactions. The corresponding FGR evaluation for such a
time-scale is explicitly performed in Ref. \cite{danieli2} and it yields:

\begin{equation}
\frac{1}{\tau }=2\frac{2\pi }{\hbar }\Delta ^{2}\frac{4}{3\pi ^{2}J}.
\label{FGR_delta}
\end{equation}%
As above, the extra $2$ factor stands for the contributions of two
semi-infinite linear chains. The factor $4/(3\pi ^{2}J)$ stands for the
corresponding LDoS evaluated at $\varepsilon =0$. Then,

\begin{equation}
\delta E=\frac{\hbar }{2\tau }=\frac{8}{3\pi }\frac{\Delta ^{2}}{J}.
\label{EnergyUncertainty_delta}
\end{equation}%
This uncertainty adds to the bandwidth and hence it leads to the dimensional
estimation of the critical line of the MBL transition,%

\begin{align}
W_{c}(\Delta )&\sim \frac{e}{2}(B+2\delta E)\notag\\
&\sim \frac{2.71}{2}\left( 2J+%
\frac{16}{3\pi }\frac{\Delta ^{2}}{J}\right) ,  \label{w_critico}
\end{align}%
which is plotted in Fig. \ref{Fig_mapa} as a dashed line starting in the
open circle.

\section{Conclusion}

We simulated the dynamics of a local Loschmidt echo in a spin system in
the presence of interactions and disorder, for a wide regime of these competing
magnitudes. The computation yields a phase diagram that evidences the
parametric region where ergodicity manifests. Non-ergodic behaviors were
classified and discussed in terms of glassy dynamics, standard Anderson
Localization and the Many-Body Localization. Based on the evaluation of
energy uncertainties introduced by weak interactions and weak disorder, we
estimated the critical lines that separate these phases. The agreement
between the estimated critical lines and the LE diagram is considerably good.

In spite of the fact that the local nature of the LE observable constitutes
a limitation to perform a reliable finite size scaling procedure, our
strategy seems promising to analyze different underlying topologies and
different ways of breaking down integrability. Last, but not least, in state-of-the-art NMR \cite{Franzoni2005,Boutis2012}, the high
temperature correlation functions, like the LE, are privileged witnesses
for the onset of phase transitions \cite{Laflamme2009LE} that could hint the
appearance of Many-Body Localization  \cite%
{Franzoni2005,Alvarez2010_localization,Alvarez2014experimental}.

\begin{acknowledgements}
PRZ and HMP wish to dedicate this paper to the memory of their coauthor
Patricia Rebeca Levstein who did not live to see the final version of this
paper. We are grateful to A. Iucci, A. D. Dente and C. Bederi\'{a}n for
their cooperation at various stages of this work. This work benefited from
fruitful discussions with T. Giamarchi and comments by F. Pastawski. We
acknowledge support from CONICET, ANPCyT, SeCyT-UNC and MinCyT-Cor. The
calculations were done on Graphical Processing Units under an NVIDIA
Professor Partnership Program led by O. Reula.
\end{acknowledgements}


\begin{thebibliography}{99}
\bibitem{LebowitzPhysToday} J~L {Lebowitz}, {\it Boltzmann's
entropy and time's arrow}, {Phys. Today} {\bf 46}, 32 (1993).

\bibitem{LebowitzRMP1999} J~L {Lebowitz}, {\it Statistical
mechanics: A selective review of two central issues}, {Rev. Mod. Phys. Supplement} {\bf 71}, 346 (1999).

\bibitem{FPU} E~Fermi, J~Pasta, S~Ulam, {\it Studies of
nonlinear problems}, {LASL Report LA1940} {\bf 5}, 977 (1955).

\bibitem{fermi_book} E Fermi, {\it Collected Pppers: United
States 1939-1954}, Vol.~2, University of Chicago Press (1965).

\bibitem{izraChaosReview} G~P Berman, F~M Izrailev, {\it
The Fermi--Pasta--Ulam problem: Fifty years of progress}, {Chaos} {\bf 15}, 015104 (2005).

\bibitem{chirikov1} B~V Chirikov, {\it Resonance processes in magnetic traps},
{J. Nucl. Energy C} {\bf 1}, 253 (1960).

\bibitem{chirikov2} F~M {Izrailev}, B~V {Chirikov}, {\it
Statistical properties of a nonlinear string}, {Sov. Phys.
Dokl.} {\bf 11}, 30 (1966).

\bibitem{Zaslavsky} G~M {Zaslavsky}, {\it Chaotic dynamics and
the origin of statistical laws}, {Phys. Today} {\bf 52}, 39 (1999).

\bibitem{LebowitzQET} S~{Goldstein}, J~L {Lebowitz}, R~{Tumulka}, N~%
{Zangh{\`i}}, 
{\it Long-time behavior of macroscopic quantum systems. Commentary
  accompanying the English translation of John von Neumann's 1929 article on
  the quantum ergodic theorem}, {Eur. Phys. J. H} {\bf 35}, 173 (2010).

\bibitem{vonNeumannQET} J~{von Neumann}, 
{\it Proof of the ergodic theorem and the H-theorem in quantum mechanics.
  Translation of: Beweis des ergodensatzes und des H-theorems in der neuen
  mechanik}, {Eur. Phys. J. H} {\bf 35}, 201 (2010).

\bibitem{jalpa} R~A Jalabert, H~M Pastawski, 
{\it Environment-independent decoherence rate in classically chaotic
  systems}, {Phys. Rev. Lett.} {\bf 86}, 2490 (2001).

\bibitem{patricia98} P~R Levstein, G Usaj, H~M Pastawski,
{\it Attenuation of polarization echoes in nuclear magnetic resonance: A
  study of the emergence of dynamical irreversibility in many-body quantum
  systems}, {J. Chem. Phys.} {\bf 108}, 2718 (1998).
  
\bibitem{scholarpedia} A Goussev, R A Jalabert, H M Pastawski, D Wisniacki, {\it Loschmidt echo}, \newblock
{Scholarpedia} {\bf 7}, 11687 (2012).

\bibitem{Zangara2012} P~R Zangara, A~D Dente, P~R Levstein, H~M  Pastawski, 
{\it Loschmidt echo as a robust decoherence quantifier for many-body
  systems}, {Phys. Rev. A} {\bf 86} 012322 (2012).

\bibitem{Jacquod} Ph Jacquod, C~Petitjean, {\it Decoherence, entanglement and irreversibility in quantum dynamical
  systems with few degrees of freedom}, {Adv. in Phys.} {\bf 58}, 67 (2009).

\bibitem{usaj-physicaA} H~M Pastawski, P~R Levstein, G~Usaj, J~Raya, J~Hirschinger, 
{\it A nuclear magnetic resonance answer to the Boltzmann-Loschmidt
  controversy?} {Physica A} {\bf 283}, 166 (2000).

\bibitem{MolPhys} G~{Usaj}, H~M {Pastawski}, P~R {Levstein}, 
{\it Gaussian to exponential crossover in the attenuation of polarization
  echoes in NMR}, {Mol. Phys.} {\bf 95}, 1229 (1998).

\bibitem{CradleNature2006} T~{Kinoshita}, T~{Wenger}, D~S {Weiss}, %
{\it A quantum Newton's cradle}, {Nature} {\bf 440}, 900 (2006).

\bibitem{BlochNatPhys2012} S~{Trotzky}, Y-A {Chen}, A~{Flesch}, I~P {McCulloch},  U~{Schollw{\"o}ck}, J~{Eisert}, I~{Bloch}, 
{\it Probing the relaxation towards equilibrium in an isolated strongly
  correlated one-dimensional Bose gas}, {Nat. Phys.} {\bf 8}, 325 (2012).

\bibitem{polkovnikovRMP} A Polkovnikov, K Sengupta, A Silva, M  Vengalattore, 
{\it Colloquium: Nonequilibrium dynamics of closed interacting
  quantum systems.} {Rev. Mod. Phys.} {\bf 83}, 863 (2011).

\bibitem{altshuler2006} D~M {Basko}, I~L {Aleiner}, B~L {Altshuler}, 
{\it Metal insulator transition in a weakly interacting many-electron
  system with localized single-particle states}, {Ann. Phys. New York} {\bf 321}, 1126 (2006).

\bibitem{Altshuler2010} I~L {Aleiner}, B~L {Altshuler}, G~V {Shlyapnikov}, 
{\it A finite-temperature phase transition for disordered weakly
  interacting bosons in one dimension}, {Nat. Phys.} {\bf 6}, 900 (2010).

\bibitem{MottRMP} N~F Mott, {\it Metal-insulator transition}, %
{Rev. Mod. Phys.} {\bf 40}, 677 (1968).

\bibitem{AndersonRMP-1978} P~W Anderson, {\it Local moments and
localized states}, {Rev. Mod. Phys.} {\bf 50}, 191 (1978).

\bibitem{popescuNATURE2006} S~{Popescu}, A~J {Short}, A~{Winter},%
{\it Entanglement and the foundations of statistical mechanics}. %
{Nat. Phys.} {\bf 2}, 754 (2006).

\bibitem{rigolNATURE2008} M~Rigol, V~Dunjko, M~Olshanii, 
{\it Thermalization and its mechanism for generic isolated quantum
  systems}, {Nature} {\bf 452}, 854 (2008).

\bibitem{oganesyanhuse2007} V Oganesyan, D~A Huse, \newblock
{\it Localization of interacting fermions at high temperature}, \newblock
{Phys. Rev. B} {\bf 75}, 155111 (2007).

\bibitem{prosen2008} M \ifmmode \v{Z}\else \v{Z}\fi{}nidari\ifmmode~\v{c}%
\else
\v{c}\fi{}, T Prosen, P  Prelov%
\ifmmode~\v{s}\else \v{s}\fi{}ek, 
{\it Many-body localization in the Heisenberg $XXZ$ magnet in a random
  field}, {Phys. Rev. B} {\bf 77}, 064426 (2008).

\bibitem{palhuse2010} A Pal, D~A Huse, {\it Many-body localization phase transition}, {Phys. Rev. B} {\bf 82}, 174411 (2010).

\bibitem{pollmann2012} J~H Bardarson, F Pollmann, J~E Moore, %
{\it Unbounded growth of entanglement in models of many-body
localization}, {Phys. Rev. Lett.} {\bf 109}, 017202 (2012).

\bibitem{DeLuca2012} A~De Luca, A~Scardicchio, {\it
Ergodicity breaking in a model showing many-body localization}, \newblock
{Europhys. Lett.} {\bf 101}, 37003 (2013).

\bibitem{Refael2013} D Pekker, G Refael, E Altman, E Demler, V Oganesyan, 
 {\it The Hilbert-glass transition: new universality of temperature-tuned
  many-body dynamical quantum criticality}, {Phys. Rev. X} {\bf 4}, 011052 (2014).

\bibitem{Giamarchi1988} T~Giamarchi, H~J Schulz, {\it
Anderson localization and interactions in one-dimensional metals}, {Phys. Rev. B} {\bf 37}, 325 (1988).

\bibitem{Fisher1992} C~A Doty, D~S Fisher, {\it
Effects of quenched disorder on spin-1/2 quantum \textit{XXZ} chains}, %
{Phys. Rev. B} {\bf 45}, 2167 (1992).

\bibitem{Kimball1981} J~Kimball, 
{\it Comments on the interplay between Anderson localisation and
  electron-electron interactions}, {J. Phys. C Solid State} {\bf 14}, L1061 (1981).

\bibitem{mesoECO-exp} H~M Pastawski, G Usaj, P~R Levstein, 
{\it Quantum interference phenomena in the local polarization dynamics of
  mesoscopic systems: an NMR observation}, {Chem. Phys. Lett.} {\bf 261}, 329 (1996).

\bibitem{madi-ernst1997} Z~L M\'adi, B~Brutscher, T~Schulte-Herbr\"uggen, R~Br\"uschweiler, R~R Ernst,
{\it Time-resolved observation of spin waves in a linear chain of nuclear
  spins}, {Chem. Phys. Lett.} {\bf 268}, 300 (1997).

\bibitem{danieli-CPL2004} E~P Danieli, H~M Pastawski, P~R Levstein, {\it Spin projection chromatography}, {Chem.
Phys. Lett.} {\bf 384}, 306 (2004).

\bibitem{MacKinnon93} B~{Kramer}, A~{MacKinnon}, {\it
Localization: Theory and experiment}, {Rep. Prog. Phys.} {\bf 56}, 1469 (1993).

\bibitem{Alv-parallelism} G~A \'Alvarez, E~P Danieli, P~R Levstein, H~M  Pastawski, 
{\it Quantum parallelism as a tool for ensemble spin dynamics
  calculations}, {Phys. Rev. Lett.} {\bf 101}, 120503 (2008).

\bibitem{Ernst1992} S Zhang, B~H Meier, R~R Ernst,
{\it Polarization echoes in NMR}, {Phys. Rev. Lett.} {\bf 69}, 2149 (1992).

\bibitem{Cappellaro2014_review} P Cappellaro, 
{\it Implementation of state transfer Hamiltonians in spin chains with
  magnetic resonance techniques}, In: {Quantum State  Transfer and Network Engineering}, Eds. G~M Nikolopoulos, I Jex, Pag. 183--222, Springer Berlin Heidelberg (2014).

\bibitem{Rhim1971} W-K Rhim, A~Pines, J~S Waugh, {\it
Time-reversal experiments in dipolar-coupled spin systems}, {Phys.
Rev. B} {\bf 3}, 684 (1971).

\bibitem{elenaMQC} E~Rufeil-Fiori, C~M S\'anchez, F~Y Oliva, H~M Pastawski, P~R  Levstein, {\it Effective one-body dynamics in multiple-quantum NMR experiments}, {Phys. Rev. A} {\bf 79}, 032324 (2009).

\bibitem{Alvarez2014experimental} G~A {\'Alvarez}, D~{Suter}, R~{Kaiser}, 
{\it Experimental observation of a phase transition in the evolution of
  many-body systems with dipolar interactions}, arXiv:1409.4562 (2014).

\bibitem{mesoECO-PRL1995} H~M Pastawski, P~R Levstein, G Usaj, {\it Quantum dynamical echoes in the spin diffusion
in mesoscopic systems}, {Phys. Rev. Lett.} {\bf 75}, 4310 (1995).

\bibitem{Fernandez2015} L~J Fern\'andez-Alc\'azar, H~M Pastawski, 
{\it Decoherent time-dependent transport beyond the Landauer-B\"uttiker
  formulation: A quantum-drift alternative to quantum jumps}, {Phys. Rev. A} {\bf 91}, 022117 (2015).

\bibitem{GangFour_PRL1979} E~Abrahams, P~W Anderson, D~C Licciardello, \raggedbottom  \pagebreak T~V Ramakrishnan, 
 {\it Scaling theory of localization: Absence of quantum diffusion in two
  dimensions}, {Phys. Rev. Lett.} {\bf 42}, 673 (1979).

\bibitem{Ziman1969} J~M Ziman, 
{\it Localization of electrons in ordered and disordered systems ii. Bound
  bands}, {J. Phys. C Solid State} {\bf 2}, 1230 (1969).

\bibitem{danieli2} E P Danieli, G A \'Alvarez, P R Levstein, H M Pastawski, 
{\it Quantum dynamical phase transition in a system with many-body
  interactions}, {Solid State Commun.} {\bf 141}, 422 (2007).

\bibitem{Franzoni2005} M~B Franzoni, P~R Levstein,
{\it Manifestations of the absence of spin diffusion in multipulse NMR
  experiments on diluted dipolar solids}, {Phys. Rev. B} {\bf 72}, 235410 (2005).

\bibitem{Boutis2012} S~W Morgan, V Oganesyan, G~S Boutis, 
{\it Multispin correlations and pseudothermalization of the transient
  density matrix in solid-state NMR: Free induction decay and magic echo},%
{Phys. Rev. B} {\bf 86}, 214410 (2012).

\bibitem{Laflamme2009LE} J Zhang, F~M Cucchietti, C~M Chandrashekar, M Laforest,  C~A Ryan, M Ditty, A Hubbard, J~K Gamble,  R  Laflamme, {\it Direct observation of
quantum criticality in Ising spin chains}, {Phys. Rev. A} {\bf 79}, 012305 (2009).

\bibitem{Alvarez2010_localization} G~A \'Alvarez, D Suter, 
{\it NMR quantum simulation of localization effects induced by
  decoherence}, {Phys. Rev. Lett.} {\bf 104}, 230403 (2010).
\end{thebibliography}

\end{document}